\documentclass[twocolumn,aps,prb,amsmath,amssymb]{revtex4-1}
\usepackage{graphicx}
\usepackage{dcolumn}
\usepackage{bm}
\usepackage{multirow}
\usepackage{float}

\begin{document}
\title{Many-body localization in a non-Hermitian quasi-periodic system}

\author{Liang-Jun Zhai$^1$, Shuai Yin$^{2}$, Guang-Yao Huang$^{3}$}
\email{guangyaohuang@quanta.org.cn}
\affiliation{$^1$School of Mathematics and Physics, Jiangsu University of Technology, Changzhou 213001, China}
\affiliation{$^2$School of Physics, Sun Yat-Sen University, Guangzhou 510275, China}
\affiliation{$^3$Institute for Quantum Information $\&$  State Key Laboratory of High Performance Computing, College of Computer, National University of Defense Technology, Changsha 410073, China}

\date{\today}

\begin{abstract}
In the present study, the interplay among interaction, topology, quasiperiodicity, and non-Hermiticity is studied.
The hard-core bosons model on a one-dimensional lattice with asymmetry hoppings and quasiperiodic onsite potentials is selected.
This model, which preserves time-reversal symmetry (TRS), will exhibit three types of phase transition: real-complex transition of eigenenergies, topological phase transition and many-body localization (MBL) phase transition.
For the real-complex transition, it is found that the imaginary parts of the eigenenergies are always suppressed by the MBL.
Moreover, by calculating the winding number, a topological phase transition can be revealed with the increase of potential amplitude, and we find that the behavior is quite different from the single-particle systems.
Based on our numerical results, we conjecture that these three types of phase transition occur at the same point in the thermodynamic limit, and the MBL transition of quasiperiodic system and disordered system should belong to different universality classes.
Finally, we demonstrate that these phase transitions can profoundly affect the dynamics of the non-Hermitian many-body system.
\end{abstract}

\maketitle

\section{\label{intro}Introduction}
In recent years, as an extension of the noninteracting Anderson localization,
a phenomenon termed as many-body localization (MBL) in the quantum many-body systems has received a lot of attention
~\cite{Abanin2019,Abanin2017,Huse2015,Rispoli2019,Kohlert2019,Morningstar2019,Lukin2019,Basko2006,Znidaric2008,Pal2010,Kjall2014,He2017,Fan2017,Gornyi2005,Panda2020}.
In such a phase, the system fails to act as a bath for its own subsystems and thermalization does not occur.
It has been established that the MBL phase has drastically different spectra and dynamical properties compared with the delocalization (thermal) phase.
Although MBL is usually studied for systems with random disorder, there is another type of system, the quasiperiodic system, which also supports MBL due to its unique features~\cite{Iyer2013,Schreiber2015,Luschen2017,Setiawan2017, Yao2018,Khemani2017,Lee2017,Mace2019,Weiner2019,Xu2019,Varma2019,Luschen20172,Zhao2019,Doggen2019,Yao2019,Goblot2020,Cookmeyer2020}.
The quasiperiodic system breaks translational invariance by the incommensurate period, and shows some randomlike properties similar to the disordered system.
However, compared with the disordered system, the quasiperiodic system has a long-range correlation, and creates a disorder in a more controlled way~\cite{Vidal2001}.
Thus the quasiperiodic system constitutes an intermediate phase between a periodic system and a fully disordered system.
In the previous theoretical studies, it has been found that MBL phase transition in the Hermitian disordered system and quasiperiodic system belongs to two distinct universality classes~\cite{Yao2018,Khemani2017},
and MBL phase in the quasiperiodic system is more stable as compared to the disordered system~\cite{Khemani2017}.

Most recently, great interest has been devoted to studying the MBL phenomena in the non-Hermitian systems~\cite{Hamazaki2019,Levi2016,Medvedyeva2016,Wu2019,Henben2020}.
The results showed that MBL signatures can be restored even in the appearance of dissipation~\cite{Levi2016}.
In some classes of non-Hermitian system with time-reversal symmetry (TRS), there is a real-complex transition of eigenenergies featuring parity-time (PT) symmetry~\cite{Ganainy2018,Ni2018,Malzard2015,Zhai2019}.
It is found that MBL can suppress the imaginary parts of the complex eigenenergies of the disordered system,
and the real-complex transition occurs accompanied with the MBL phase transition~\cite{Hamazaki2019}.
On the other hand, exotic topological phases were unveiled in the non-Hermitian quantum systems~\cite{Yao20181,Yao20182,Zeuner2015,Shen2018,Jin2018,Leykam2017,Kawabata2019,Gong2018,Longhi2019,Longhi20192,Liu2019,Zhang2020,Zeng2020,Yoshida2019,Liu2020}.
For the non-Hermitian single particle systems, theoretical studies found that the localization-delocalization phase transition for both the disordered and quasiperiodic systems has a topological nature,
and the localization and delocalization phases can be characterized by the winding number~\cite{Gong2018,Longhi2019,Zhang2020}.
For the Hermitian many-body systems, it has been found that the interactions can destroy the topological phases or create new topological phases which are topologically distinct from the trivial states~\cite{Chiu2016,Lapa2015,Maciejko2010,Shackleton2020},
and MBL eigenstates can exhibit or fail to exhibit topological orders~\cite{Huse2013,Chandran2014,Kuno2019,Orito2019}.
However, for the non-Hermitian many-body systems, there are few works that have been done to investigate the affection of the interaction on the topological phase and the relations between the topological and MBL phase transitions.

With this background, the interplay among interaction, topology, quasiperiodicity, and non-Hermiticity is examined in this paper.
The study is applied to a hard-core bosons model on a one-dimensional lattice with asymmetry hoppings and quasiperiodic onsite potentials.
The non-Hermiticity of the model comes from asymmetry hopping, but it still has the TRS.
We find that the MBL phase transition, real-complex transition and topological phase transition coexist for this model,
and the transition points of these transitions are close.
The obtained critical exponent of quasiperiodic system is different from the disordered system, which means that they belong to different universality classes.
Based on our numerical results, we conjecture that these three phase transitions occur at the same point in the thermodynamic limit.
Since the real-complex transition and MBL phase transition can profoundly affect the dynamics of the non-Hermitian system~\cite{Yin2019},
the dynamical behaviors of the real part of eigenenergy and entanglement entropy are studied.

The remainder of the paper is organized as follows. In Sec.~\ref{secmodel}, the model of the non-Hermitian quasiperiodic system is presented. Numerical investigation is presented in Sec.~\ref{secResult}.
A summary is given in Sec.~\ref{secSum}.

\section{\label{secmodel}Model}
A non-Hermitian hard-core bosons model on a one-dimensional lattice is considered in the present study.
The Hamiltonian reads
\begin{eqnarray}
\hat{H}=&&\sum_{i=1}^{L}{[-J(e^{-g}\hat{b}_{i+1}^{\dag}\hat {b}_{i}+e^{g}\hat{b}_{i}^{\dag}\hat {b}_{i+1})+U \hat{n}_{i}\hat{n}_{i+1}}\label{model}\\ \nonumber
  &&+W_i\hat{n}_i].
  \end{eqnarray}
Here, $\hat {b}_{i}$ and $\hat {b}_{i}^\dag$ are the annihilation and creation operators of a hard-core boson, and $\hat{n}_i=\hat{b}_{i}^{\dag}\hat {b}_{i}$ is the particle-number operator at site $i$.
$J$ and $g$ label the asymmetry hopping amplitude between the nearest-neighboring (NN) sites, and $U$ is the interaction between NN sites.
For a quasiperiodic system, the onsite potential is $W_i=W\cos{(2\pi\alpha i+\phi)}$, where $W$ is the amplitude of the potential, and $\phi$ is the phase of the potential, and $\alpha$ is irrational for incommensurate potentials.

For this model, the non-Hermiticity is controlled by the parameter $g$, but it still has the TRS.
In the following, we assume $J=1$, $U=2$, $g=0.5$, and the subspace with fixed particle number $M=L/2$ will be selected.
The irrational number $\alpha$ is chosen as the inverse of the golden ratio $\alpha=(\sqrt{5}-1)/2$, which could be compared with the experimental results~\cite{Schreiber2015}.
The periodic boundary condition is assumed in the following calculation.

\section{\label{secResult}Numerical results and discussions}
\subsection{Real-complex transition of eigenenergies}
\begin{figure}
  \centering
  \includegraphics[width=3 in]{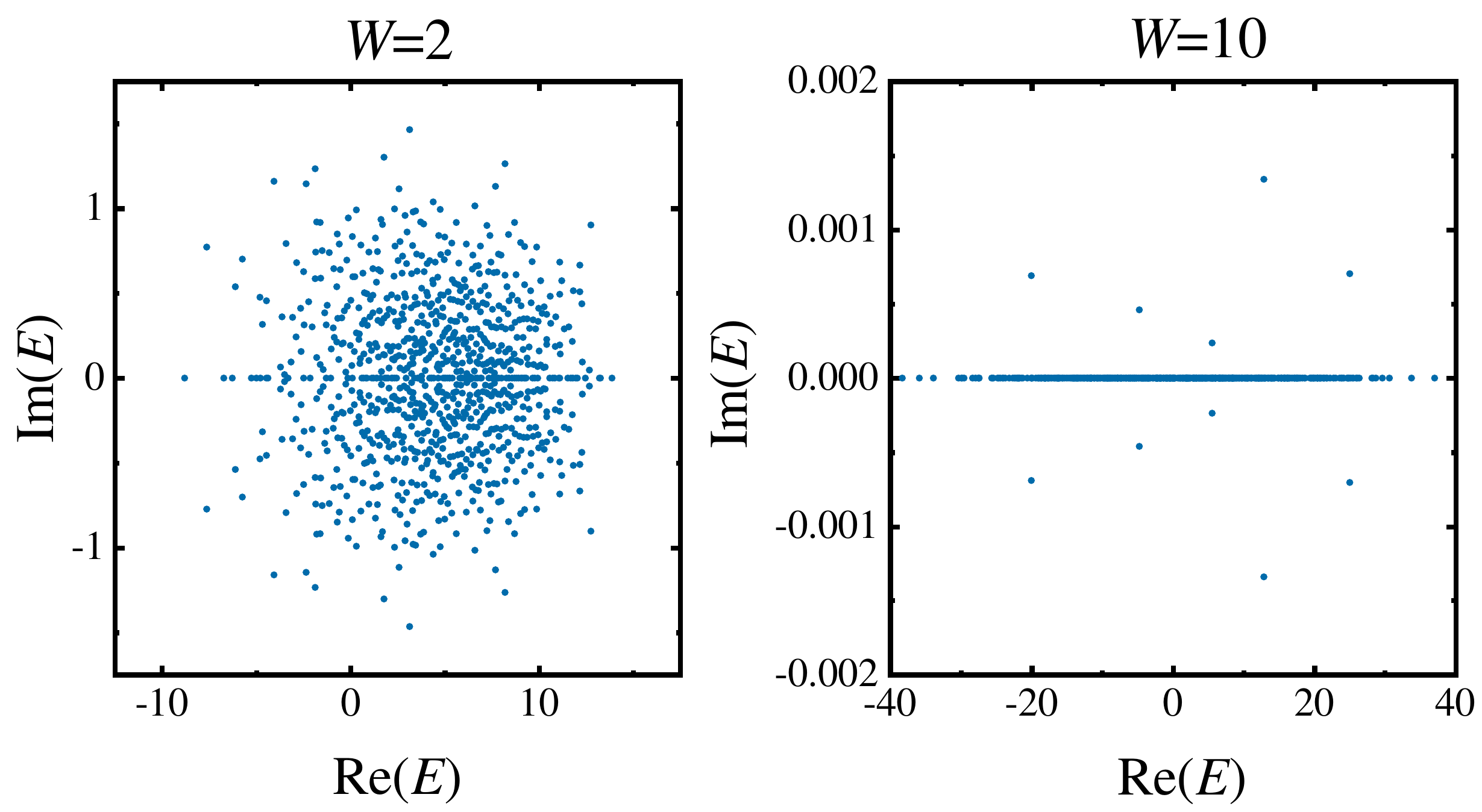}\\
  \caption{Eigenenergies of the Hamiltonian Eq.~(\ref{model}) with $W=2$ (left) and
  $W=10$ (right). Here, the lattice size is $L=12$.}
  \label{fig:ImE}
\end{figure}

Firstly, the real-complex transition of this model is studied.
As shown in Fig.~\ref{fig:ImE}, the eigenenergies of Hamiltonian Eq.~(\ref{model}) with $L=12$ and different $W$ are plotted.
Since the non-Hermitian Hamiltonian still has the TRS, it is found that the imaginary parts of the spectra are symmetric around the real axis.
With the increase of $W$, the eigenenergies with nonzero imaginary part decrease.

\begin{figure}
  \centering
  \includegraphics[width=3 in]{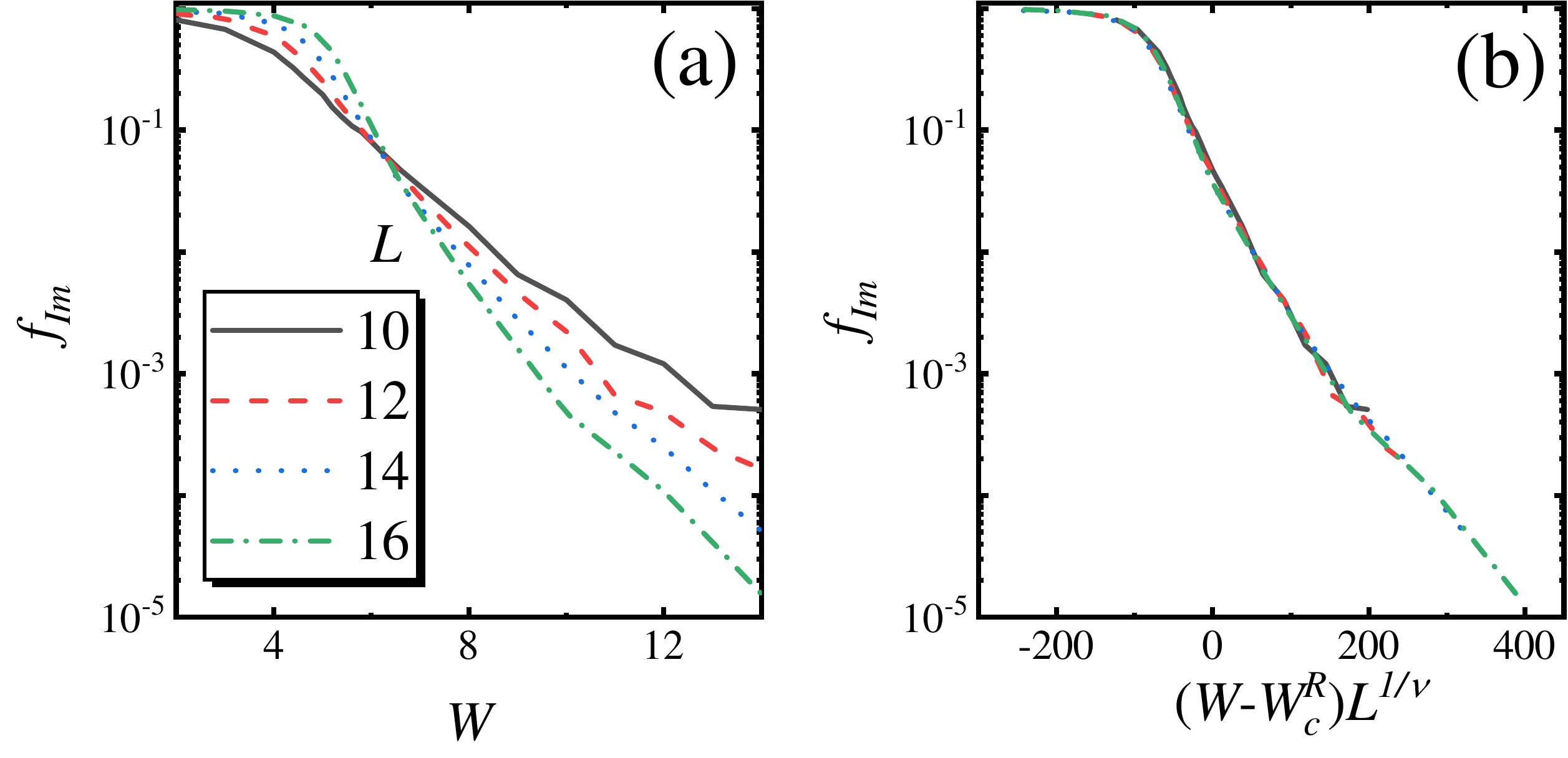}\\
  \caption{(a) $f_{Im}$ as a function of $W$ for $L=10, 12, 14$ and 16, and (b) the rescaled curves according to Eq.~(\ref{real-complex}).}
  \label{fig:fIm}
\end{figure}

To measure the variation of the ratio of the complex eigenenergies with nonzero imaginary part, $f_{Im}$ is defined as
\begin{eqnarray}
 f_{Im}=\overline{D_{Im}/D},
\end{eqnarray}
where the $D_{Im}$ is the number of eigenenergies with nonzero imaginary part, and $D$ is the total number of eigenenergies.
Here, a cutoff of $C=10^{-13}$ is used, that is, $|\mathrm{Im} E|\leq C$ is identified to be a machine error.
In Fig.~\ref{fig:fIm}(a), $f_{Im}$ as a function of $W$ for different $L$ is plotted, and the results are obtained by averaging 500 choices of $\phi$ for $L=10, 12$, and 14, and 100 choices of $\phi$ for $L=16$.
Roughly speaking, when $W\leq W_C^{\rm R}=6.6$, $f_{Im}$ increases with the increase of $L$, while $f_{Im}$ decreases with the increase of $L$ for $W\geq W_C^R$.
The curves of $f_{Im}$ versus $W$ can be rescaled by the following scaling function
\begin{eqnarray}
  f_{Im} &\propto& (W-W_C^{\rm R})L^{1/\nu},
  \label{real-complex}
\end{eqnarray}
where $\nu=0.7$.
As shown in Fig.~\ref{fig:fIm}(b), the rescaled curves collapse onto each other, which confirms Eq.~(\ref{real-complex}).
These results demonstrate that in the thermodynamic limit ($L\rightarrow\infty$) the model of Eq.~(\ref{model}) should have a real-complex transition at $W=W_C^{\rm R}$, that is, when $W<W_C^{\rm R}$ the eigenenergies are almost complex, while the eigenenergies are almost real for $W>W_C^{\rm R}$.
Similar results have also been found in the non-Hermitian Hamiltonian with a random onsite potential~\cite{Hamazaki2019}, but the scaling exponent $\nu$ for the disordered system is different from what we found here, which implies the real-complex transition should be in different universality classes for the disordered and quasiperiodic systems.

\begin{figure}
  \centering
  \includegraphics[width=3 in]{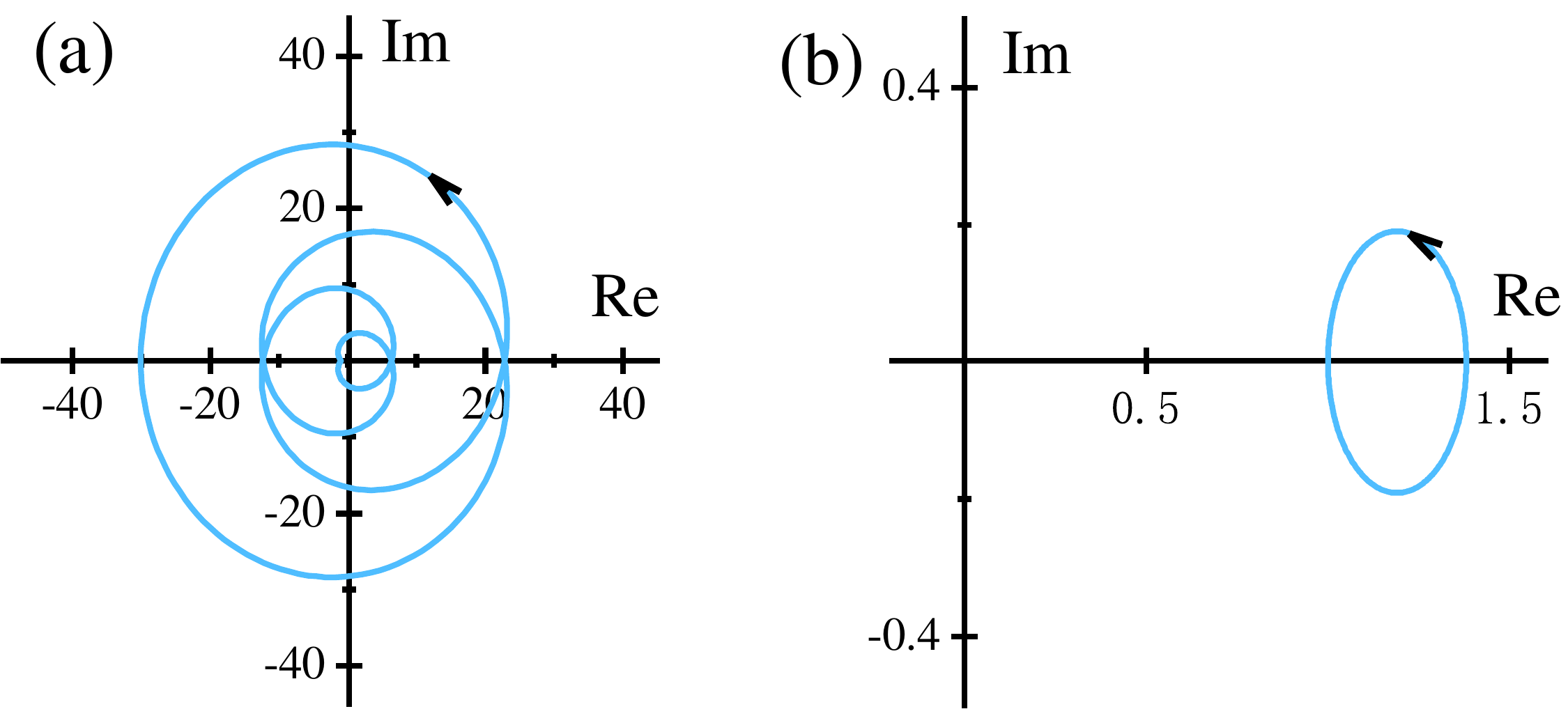}\\
  \caption{(a) The $\Phi$ dependence of $\det H(\Phi)/|\det H(0)|$ in the complex plane for (a) $W=3.5$ and (b) $W=6.5$.
  The lattice size is $L=10$ and $\phi=\pi/6$.
  }\label{fig:DetH}
\end{figure}

\subsection{Topological phase transition}

\begin{figure}
  \centering
  \includegraphics[width=3 in]{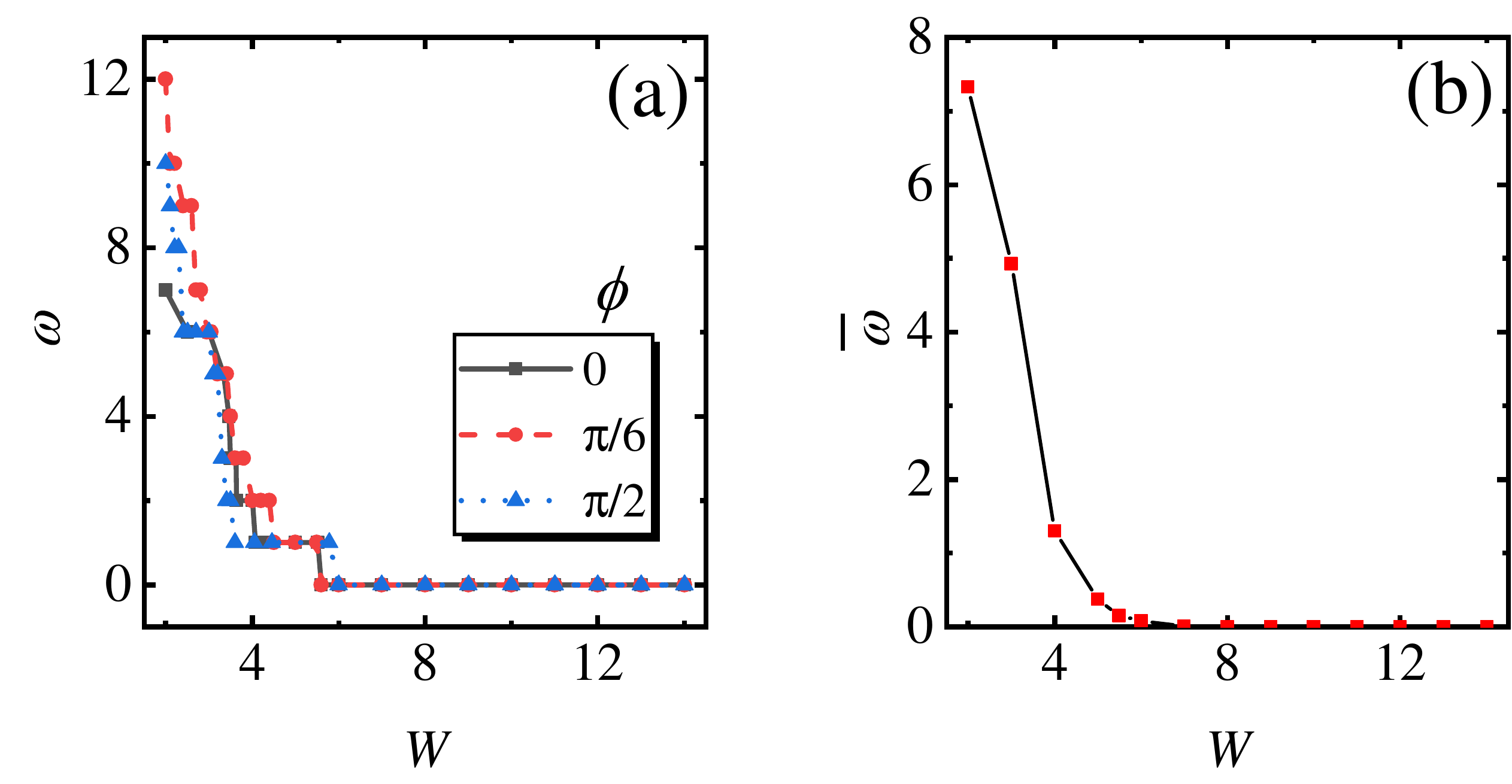}\\
  \caption{(a)The $W$ dependence of winding number $\omega$ with $\phi=0, \pi/6$ and $\pi/2$, (b) and the $W$ dependence of averaged winding number $\overline{\omega}$.
  The lattice size is $L=10$.
  }\label{fig:Winding}
\end{figure}
Different from the Hermitian systems, to study the topological phases of non-Hermitian systems, not only the ground state but also the full complex spectra should be taken into account~\cite{Gong2018,Longhi2019}.
Therefore, a natural topological object arising from the complex energy plane is the winding number, that is, a loop constituted by eigenenergies which encircles a prescribed base point.
The winding number is topologically stable and changes its value only when the curve is crossing the base point.
Recently, the winding number has been defined to study the topological phase for non-Hermitian systems in the single particle picture without interactions~\cite{Gong2018,Longhi2019}.
Generalizing the idea of defining the winding number to our interacting non-Hermitian systems, a parameter $\Phi$ is introduced through a gauge transformation $\hat {b}_{j} \to e^{i\frac{\Phi}{L}j}\hat {b}_{j}$ and $\hat {b}_{j}^{\dag} \to e^{-i\frac{\Phi}{L}j}\hat {b}_{j}^{\dag}$, which can be viewed as a magnetic flux $\Phi$ through non-Hermitian ring with length $L$ is applied. The Hamiltonian becomes
 \begin{eqnarray}
   H(\Phi) &=& \sum_{j=1}^{L}{[-J(e^{-g}e^{-i\frac{\Phi}{L}}\hat {b}_{j+1}^{\dag}\hat {b}_{j}+e^{g}e^{i\frac{\Phi}{L}}\hat {b}_{j}^{\dag}\hat {b}_{j+1})}\\ \nonumber
   && +U \hat{n}_{j}\hat{n}_{j+1}+W_i\hat{n}_j],
   \label{HamiltonianWN}
 \end{eqnarray}
and subsequently the winding number is defined as~\cite{Gong2018}
\begin{eqnarray}
  \omega &=& \int_{0}^{2\pi}{\frac{d\Phi}{2\pi i}\partial_{\Phi}\ln \det\{ H(\Phi)-E_B\}}.
\end{eqnarray}
Here, $E_B$ is the prescribed basis point which is not an eigenenergy of $H(\Phi)$.
Different from the bulk-edge correspondence in the Hermitian systems, a positive (negative) winding number $\omega$ implies a $\omega$ ($-\omega$) independent edge modes localized at the left (right) boundary in the semi-infinite space.
As demonstrated in Ref.~\cite{Gong2018}, the winding number does not depend on $E_B$.
The basis point is chosen as $E_B=0$ in the calculation, so that the loop ensures the coexistence
of the $\mathrm{Im}E<0$ and $\mathrm{Im}E>0$.

It is not convenient to directly show the loop of eigenenergies for the many-body systems, alternatively,
we here use the $\Phi$ dependence of $\det H(\Phi)/|\det H(0)|$ to illustrate the loop winding around the base point~\cite{Arikawa2010}.
During the variation of $\Phi$ from $0$ to $2\pi$, $\det H(\Phi)/|\det H(0)|$ draws a closed loop in the complex plane, and if the loop winds around the origin $m$ times, the winding number is $\pm m$ ($+$ means a counterclockwise winding, while $-$ means a clockwise winding).
In Fig.~\ref{fig:DetH}, the $\Phi$ dependence of $\det H(\Phi)/|\det H(0)|$ with different $W$ for $L=10$ is plotted, and the phase is chosen as $\phi=\pi/6$.
As seen in Fig.~\ref{fig:DetH} (a), $\det H(\Phi)/|\det H(0)|$ draws a closed curve with surrounding the origin four times in the complex plane for $W=3.5$, while $\det H(\Phi)/|\det H(0)|$ draws a closed curve without surrounding the origin for $W=6.5$ shown in Fig.~\ref{fig:DetH} (b).
It gives that $\omega=4$ for $W=3.5$ and $\omega=0$ for $W=6.5$.

Since the irrational period breaks the translational invariance, the energy spectra change with $\phi$.
Therefore, for a specific $W$, the winding number $\omega$ also changes with the phase $\phi$.
In Fig.~\ref{fig:Winding} (a), the $W$ dependence of $\omega$ with different $\phi$ for $L=10$ is plotted.
Although $\phi$ induces some differences for these curves, some behaviors are in common.
On the one hand, $\omega$ decreases with an increase of $W$, which is different from the single particle non-Hermitian system.
For the single particle non-Hermitian system, the winding number is found as $\omega=\pm 1$ for the topological phase, which means the many-body non-Hermitian systems have more complicated topological phases.
On the other hand, different curves show that a transition from topological phases with $\omega>0$ to the trivial phase ($\omega=0$) appears around $W=7$.
The average winding number $\overline{\omega}$ is plotted in Fig.~\ref{fig:Winding} (b), and it is shown that the topological phase transition point is around $W_C^{\rm T}=7$.

It should be noted that the topological transition is not equal to the disappearance of the imaginary part of eigenenergies, although the imaginary parts of eigenenergies are necessary to construct a close loop in the energy plane.
The topological phase transition gives another viewpoint on the MBL energy spectra complemented to the real-complex transition.
This winding number, defined in the complex plane by the gauge transformation, serves as a collective indicator of the eigenenergies being complex or real of the original Hamiltonian.

\subsection{MBL phase transition}

\begin{figure}
  \centering
  \includegraphics[width=3 in]{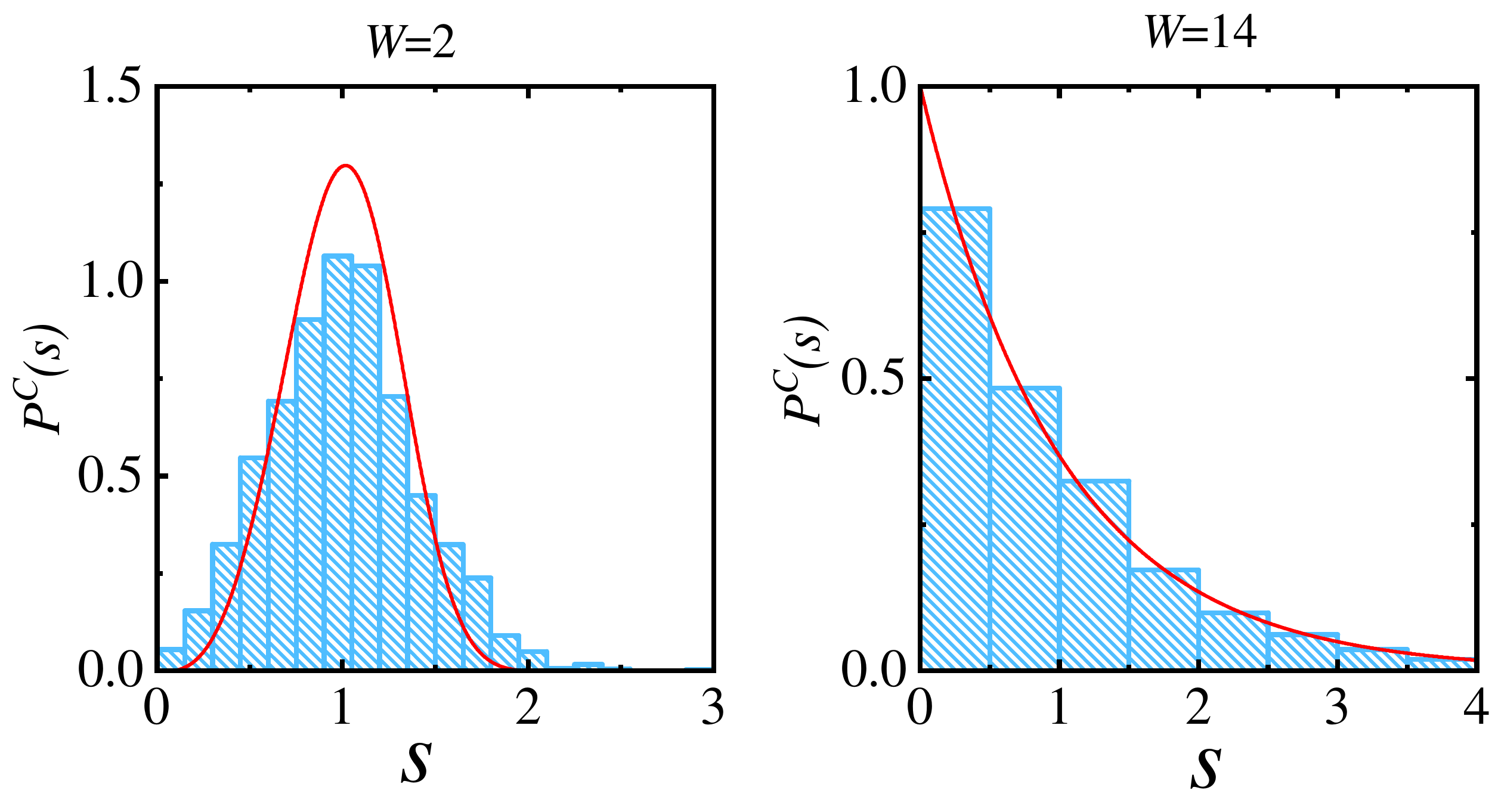}\\
  \caption{(a) The nearest-level-spacing distribution (unfolded) for $W=2$ (left) and $W=14$ (right).
  The lattice size is 16, and $\phi=\pi/4$.}\label{fig:PCS}
\end{figure}

To characterize the MBL in the non-Hermitian systems, the nearest-level-spacing distribution of eigenenergies has been generalized from the Hermitian systems~\cite{Hamazaki2019}.
On the complex plane, the nearest-level spacings for an eigenenergy $E_a$ (before unfolding) are defined as
the minimum distance of $|E_a-E_b|$.
For the delocalization phase, it has been demonstrated that the statistics of the nearest-level spacing obey a Ginibre distribution $P_{Gin}^c(s)=cp(cs)$, where
\begin{eqnarray}
p(s)=\lim_{N\rightarrow\infty}\left[\prod_{n=1}^{N-1}e_n(s^2)e^{-s^2}\right]\sum_{i=1}^{N-1}{\frac{2s^{2n+1}}{n!e_n(s^2)}},
\end{eqnarray}
with $e_n(x)=\sum_{m=0}^{n}{\frac{x^m}{m!}}$ and $c=\int_{0}^{\infty}sp(s)ds=1.1429$~\cite{Markum1999,Haakebook}.
Since the MBL tends to suppress the imaginary part of the complex eigenenergies, these eigenenergies are almost real in the MBL phase, and the nearest-level-spacing distribution becomes the Poissonian as $P_{Po}^R(s)=e^{-s}$.
By taking the eigenenergies lying within $\pm10\%$ of the real and imaginary parts from the middle of the spectra of Eq.~(\ref{model}), the nearest-level-spacing distributions (unfolded) for different $W$ are plotted in Fig~.\ref{fig:PCS}.
It is shown that for $W=2$ the distribution is a Ginibre distribution and the distribution is a Poisson distribution for $W=14$.
These results demonstrate that the non-Hermitian quasiperiodic system also has a MBL phase transition with the increase of $W$.

\begin{figure}
  \centering
  \includegraphics[width=2.2 in]{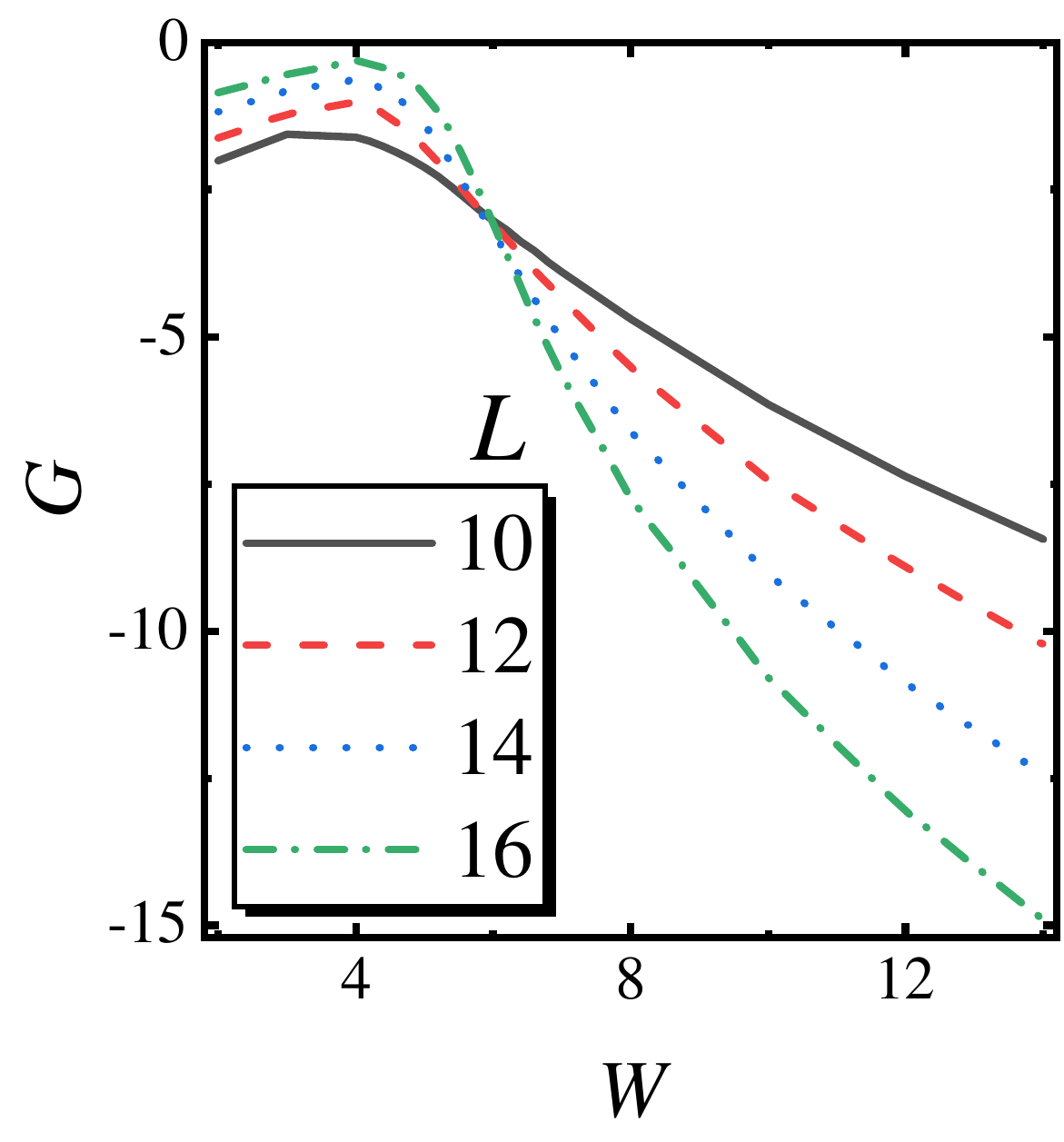}\\
  \caption{${G}$ as a function of $W$ for $L=10, 12, 14$, and 16.
  The results are obtained by averaging 500 choices of $\phi$ for $L=10, 12$ and 14, and 100 choices of $\phi$ for $L=16$.}\label{fig:QSG}
\end{figure}

\begin{figure}
  \centering
  \includegraphics[width=2.1 in]{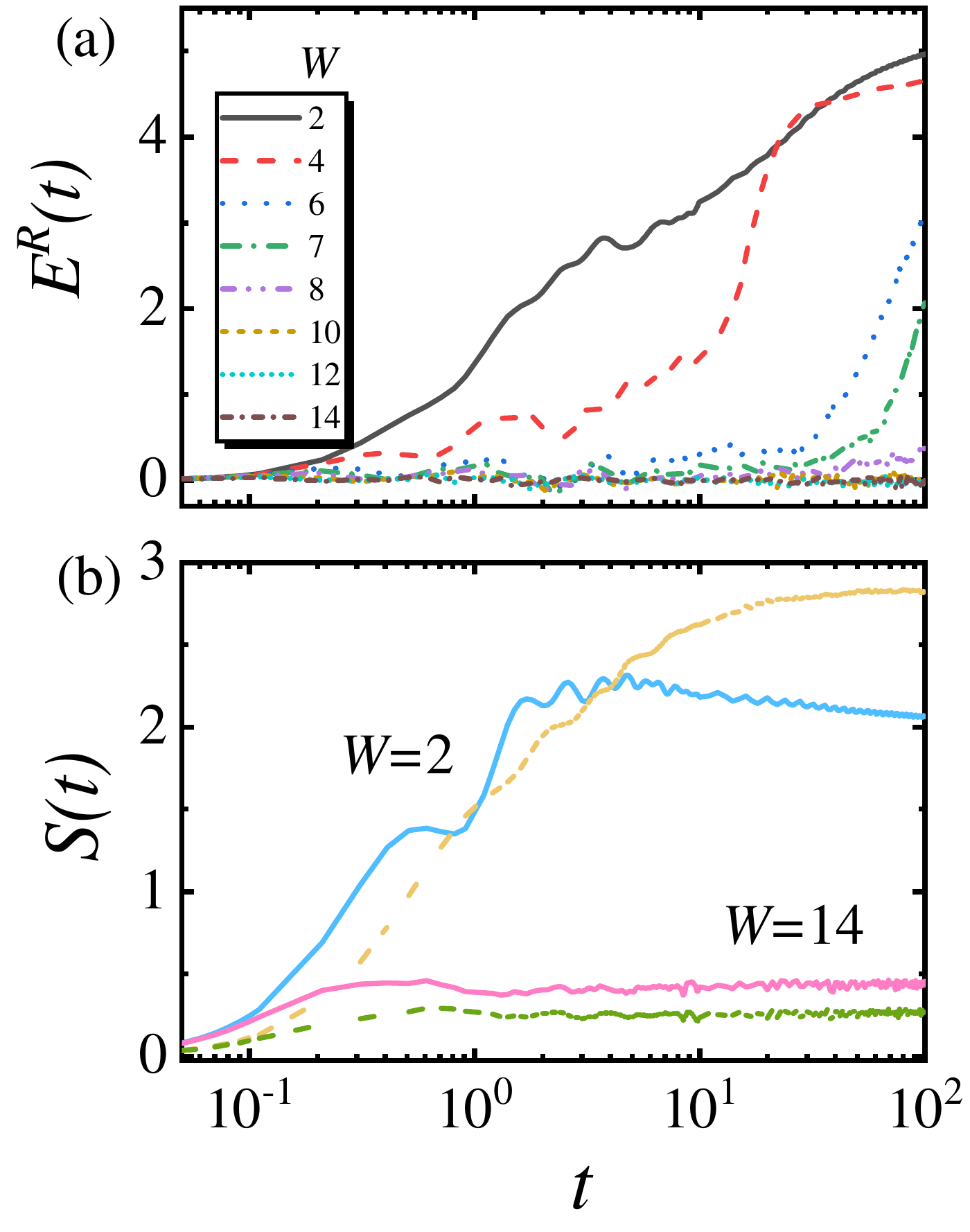}
  \caption{Time evolution of $E^R(t)$ for $W=2, 4, 6, 7, 8, 10, 12$, and 14 (a) and the time evolution of $S(t)$ for $g=0.5$ (solid lines) and $g=0$ (dotted lines) (b).
  The lattice size is $L=12$, and the initial state is taken as $|\psi_0\rangle=|1010\cdots\rangle$.}\label{fig:Dyn}
\end{figure}
Based on the response of the system's eigenstates to a local perturbation, a dimensionless parameter ${G}$ has been introduced to detect the MBL phases transition in the Hermitian systems~\cite{Serbyn2015}.
It is shown that ${G}$ decreases with the increase of system size $L$ in the MBL phase and grows with the increase of system size $L$ in the many-body delocalization phase for the Hermitian systems, and the phase transition point appears when $G(L)$ is independent of $L$.
For non-Hermitian system, since the stability of the eigenstates under perturbations $\hat{V}$ is also important for the complex eigenenergies, $G(L)$ has been extended to study the MBL phase transition in non-Hermitian systems~\cite{Hamazaki2019}.
${G}(L)$ for the non-Hermitian systems is defined as~\cite{Hamazaki2019}
\begin{eqnarray}
  G &=& \overline{\ln\frac{\langle \psi^l_{\alpha+1}|\hat{V}|\psi^r_{\alpha}\rangle}{|E'_{\alpha+1}-E'_{\alpha}|}},
\end{eqnarray}
where $\langle\psi^l_{\alpha}|$ and $|\psi^r_{\alpha}\rangle$ are the left and right eigenvectors of non-Hermitian system, $\hat{V}$ is a perturbation operator,
and $E'_{\alpha}=E_{\alpha}+\langle\psi^l_{\alpha}|\hat{V}|\psi^r_{\alpha}\rangle$ is the modified eigenenergy.
Here, the states with $E'_{\alpha}$ stays real are only considered, and the perturbation operator is selected as $\hat{V}=\hat{b}_{i+1}^+\hat{b}_i$.
In Fig.~\ref{fig:QSG}, ${G}$ as a function of $W$ for different $L$ is plotted.
It is found that in the many-body delocalization phase the absolute value of ${G}$ decreases with increase of $L$, while the absolute value of ${G}$ grows with the increase of $L$ in the MBL phases, and the MBL phase transition occurs at $W_C^{\rm MBL}=6\pm 0.2$.

From our numerical calculation, we find that $W_C^{\rm R}$, $W_C^{\rm T}$, and $W_C^{\rm MBL}$ are close,
and the slight difference is ascribed to the finite-size effect.
Therefore, based on the numerical results, we can conjecture these transition points should coincide in the thermodynamic limit.

\subsection{Effects on the dynamical behaviors}
Finally, the effects of the phase transition on the dynamical stability of the non-Hermitian systems are studied.
To illustrate this, the time evolution of real part of energy $E^R(t)$ and half-chain entanglement $S(t)$ are studied.
$E^R(t)$ is defined as
\begin{eqnarray}
E^R(t)=\mathrm{Re}\overline{[\langle\psi^r(t)|\hat{H}|\psi^r(t)\rangle]}.
\end{eqnarray}
Here, $\langle\psi^r(t)|$ is the Hermitian conjugate of $|\psi^r(t)\rangle$.
As displayed in Fig.~\ref{fig:Dyn} (a), the time evolution of $E^R(t)$ with different $W$ for $L=12$ are plotted.
It is found that for $W<W_C^{\rm R}$ and around $W_C^{\rm R}$, $E^R(t)$ changes significantly during the evolution since the nonzero imaginary part of the eigenenergies can induce dynamical instability.
For $W$ is much larger than $W_C^{\rm R}$, $E^R(t)$ is almost conserved during the evolution, since the eigenvalues are almost real.

The half-chain entanglement $S(t)$ is evaluated by the von Neumann entropy,
\begin{eqnarray}
  S(t) &=& - \overline{\mathrm{Tr}(\rho(t)\ln\rho(t))},
\end{eqnarray}
where $\rho(t)=\mathrm{Tr}_{L/2}[|\psi^r(t)\rangle\langle \psi^r(t)|]/\langle \psi^r(t)|\psi^r(t)\rangle$ is the reduced density matrix of the right eigenstate.
The time evolution of $S(t)$ for different $W$ are plotted in Fig.~\ref{fig:Dyn}(b), and the Hermitian case of $g=0$ is also shown for a comparison.
For $W=14$, the dynamical behaviors of $S(t)$ for $g=0.5$ are similar to that of $g=0$, since the eigenvalues are almost real for $g=0.5$.
However, for $W=2$, $S(t)$ first linearly grows for both values of $g$, but decreases for $t\simeq5$ only for $g=0.5$.
In addition, the long-time value of the entanglement of the many-body delocalization phase is larger than that of the MBL phase.
The reason is that entanglement entropy in the MBL phase still obeys the area law rather than the volume law even in the non-Hermitian system~\cite{Bauer2013}.

\section{\label{secSum}Summary}
In this paper, we have studied the real-complex transition, topological phase transition
and MBL phase transition in a non-Hermitian quasiperiodic system having the TRS.
Our numerical results showed that these three types of phase transitions coexist for this model, and in the thermodynamic limit these three transition points should coincide.
These results demonstrated that the imaginary parts of the eigenenergies are always suppressed by the MBL, and the MBL phase transition should have a topological nature similar to that of the single particle systems.
Moreover, the obtained critical exponent for the real-complex transition is different from that of the disordered system, which means the non-Hermitian many-body disordered system and the many-body quasiperiodic system should belong to different universality classes.
Finally, we find that these phase transitions can affect the dynamical stability, but the dynamical entanglement still obeys the area law for the MBL phase and volume law for the delocalized phase.
Recently, the asymmetry hopping has been realized experimentally in an ultracold atomic system~\cite{Gong2018}, and the MBL in the quasiperiodic system has also been experimentally studied~\cite{Rispoli2019}.
Therefore, we expect our study can be verified in these experiments.
The diagonal energy shifts of non-Hermitian systems were demonstrated to have a power-law distribution due to non-orthogonality of right and left eigenvectors~\cite{Fyodorov2018}, which is different from the nearest-level-spacing distribution. It should be an interesting work to study the diagonal energy shifts in the non-Hermitian MBL and many-body delocalization phases. We leave this for further studies.

\section*{Acknowledgments}
L.J.Z. is supported by the Natural Science Foundation of Jiangsu Province, China (Grant No. BK20170309)
and National Natural science Foundation of China (Grant No. 11704161).
S.Y. is supported by the startup grant (No. 74130-18841229) at Sun Yat-Sen University.

\end{document}